\documentclass[useAMS, usenatbib]{mn2e}
    
\usepackage{graphicx}
\usepackage{aas_macros}
\usepackage{amsmath, amssymb}
\usepackage{array}
\usepackage[usenames,dvipsnames]{color}
\usepackage{multirow}
\usepackage{threeparttable}
\usepackage{comment}
\usepackage{float}

\newcommand{\myemail}{Mohammad.Nawaz@anu.edu.au}

\usepackage{xparse}
\usepackage{soul}
\usepackage[usenames,dvipsnames]{xcolor}

\makeatletter
\NewDocumentCommand{\sayw}{O{Green}O{Black}+m}
    {%
        \begingroup
        \setulcolor{#1}%
        \setul{-0.6ex}{1.4pt}%
        \def\SOUL@uleverysyllable{%
            \rlap{%
                \color{#2}\the\SOUL@syllable
                \SOUL@setkern\SOUL@charkern}%
            \SOUL@ulunderline{%
                \phantom{\the\SOUL@syllable}}%
        }%
        \ul{#3}%
        \endgroup
    }
\makeatother

\title[Jet-ICM interaction of Hydra A]{Jet-Intracluster Medium interaction in Hydra A. II The Effect of Jet Precession}
\author[M. A. Nawaz et al.]{
M.~A. Nawaz$^{1}$,\thanks{E-mail Mohammad.Nawaz@anu.edu.au} 
G.~V. Bicknell$^{1}$,
A.~Y. Wagner$^{2}$,
R.~S. Sutherland$^{1}$ 
and B.~R. McNamara$^{3}$
\\
$^{1}$ Research School of Astronomy and Astrophysics, The Australian National University, ACT 2611, Australia; \myemail \\
$^{2}$ Center for Computational Science, Tsukuba University, 1-1-1 Tennodai, Tsukuba, Ibaraki, 305-8577 Japan \\
$^{3}$ Department of Physics and Astronomy, University of Waterloo, Waterloo, ON N2L 2Y5, Canada}

\begin{document}


\pagerange{\pageref{firstpage}--\pageref{lastpage}} \pubyear{2015}

\maketitle

\label{firstpage}

\begin{abstract}
We present three dimensional relativistic hydrodynamical simulations of a precessing jet interacting with the intracluster medium and compare the simulated jet structure with the observed structure of the Hydra A northern jet. For the simulations, we use jet parameters obtained in the parameter space study of the first paper in this series and probe different values for the precession period and precession angle. We find that for a precession period $P\approx 1 \> \rm Myr$ and a precession angle $\psi \approx 20^{\circ}$ the model reproduces i) the curvature of the jet, ii) the correct number of bright knots within 20~kpc at approximately correct locations, and iii) the turbulent transition of the jet to a plume. 
The Mach number of the advancing bow shock $\approx 1.85$ is indicative of gentle cluster atmosphere heating during the early stages of the AGN's activity. 
\end{abstract}


%
%
\section{Introduction}
It is generally agreed that the power delivered by AGN jets in massive cluster galaxies offsets the X-ray cooling of the hot cluster atmospheres \citep{birzan04a,rafferty06a,mcnamara07a,cavagnolo10a,fabian12a,mcnamara12a}. These jets occur in the majority of cluster central galaxies, in particular if they exhibit a cool core \citep{mittal09}.
Nevertheless, there are differing views as to how significant the jet power contribution is in the case of high mass clusters, based on an estimated decline of jet power with either cooling luminosity, radio luminosity, or mass \citep{best07a,turner15a}. Recent papers by \citet{hlabacek15} 
and \citet{main15a}, which include new data, support the view that the radio jet provides sufficient heating power, even in high mass clusters, whilst acknowledging the uncertainties in the slopes of the various relationships of jet power with X-ray and radio luminosities and mass. Moreover, in the particular case of Hydra~A, which we are considering here, we confirmed in \citet{nawaz14a} (hereafter Paper I) the \citet{wise07a} result, that the jets certainly have enough power ($\sim 2 \times 10^{45} \> \rm ergs \> s^{-1}$) to counteract the cooling X-ray luminosity of $4 \times 10^{44} \> \rm ergs \> s^{-1}$ \citep{david90a}. 

It is also believed that AGN jet feedback is responsible for shaping the bright end of the galaxy luminosity function \citep{croton06, fabian12, mcnamara12} through the prevention of cooling flows into the central regions of clusters.  However, the details of the energy transfer from relativistic jet plasma to the internal energy of the thermal cluster gas are not well understood. Some combination of shock heating, entrainment, thermal conduction, and magnetohydrodynamic turbulence may be involved, but the relative importance of these processes is unknown. Hence, it is important to model thoroughly well-observed sources exhibiting radio-mode feedback, such as Hydra A, in order to pin down the heating mechanisms in AGN feedback.

Hydrodynamic models of jet-ICM interactions studying the effect of AGN jets on the cooling flow atmosphere, address the long term balance between heating and cooling \citep{vernaleo06, gaspari13a}, but they also encounter difficulties. Using 3-dimensional hydrodynamical models, \cite{vernaleo06} showed that straight jets advancing into static atmospheres punch through the gas, with transient heating of the inner core of the galaxy atmosphere but with much of the mechanical energy being deposited at large radii, as the radio source evolves. As a result, straight jets are inefficient at inhibiting cooling flows. Instead, the precession of the jets and atmospheric motions driven by merging would enhance the coupling between the jet and ambient gas. For example, \citet{heinz06} showed with hydrodynamic simulations that bulk intracluster gas motions allow the jet to impact a larger volume of gas including cooled gas that would be out of the jet's reach in a static cluster atmosphere, and allow for a stronger interaction by collapsing the channels formed by the jets. In this paper, we confirm that a precessing jet distributes its kinetic energy and momentum in a wide volume extending beyond the precession cone resulting in a gentle heating of the atmosphere and a complex jet-lobe morphology.

Morphologically, extragalactic radio sources have either straight jets or complex curved morphologies with C or S shaped\footnote{this is sometimes referred to as X or Z symmetry} symmetry \citep{zaninetti88}. In general, C-symmetric jets are the result of the motion of the host galaxy with respect to the intergalactic medium \citep{douglass08, morsony13}. However, for S-symmetry there are three possible explanations: -- jet deflection by buoyancy \citep{kraft05}, jet deflection by back flows \citep{hodges-kluck11} and jet precession \citep{kurosawa08}.

In many cases jet precession is an attractive interpretation for an S-symmetric structure. The notion of jet precession was first introduced by \citet{ekers78a} who interpreted the S-shaped structure of NGC~326 as a result of the precessional motion of the jets. Subsequently,  \citet{gower82} showed, with an analytical model, that the curved jet morphologies of a number of radio galaxies may be attributed to jet precession.

Several attempts have been made to model the interaction between a precessing jet and the ambient medium numerically. Using three dimensional hydrodynamical simulations, \citet{cox91} showed that multiple hotspots of jets in radio sources are produced when the jets change their direction as a result of precessional motion. \citet{hardee01} computed 3D models of a precessing cylindrical jet and discussed the jet knots as a result of the wave-wave interactions of the body mode and surface mode of the Kelvin-Helmholtz (KH) instability. They applied their model to the inner knots of M87. \citet{kurosawa08} modelled a precessing jet originating from a precessing accretion disk with a range of precession periods and precession angles. They showed that jet precession is able to produce S- or Z-shaped structures. \citet{falceta10a} also modelled the radio source Perseus~A using three-dimensional precessing jet simulations, deriving a precession period $\approx 5 \times 10^7 \> \rm yr$.

In this paper, we show that the internal 20~kpc structure of Hydra A jet can also be modelled by a precessing jet. Based on a parameter space study we estimate the precession period and precession angle.
 
The radio galaxy Hydra A, located at the centre of the galaxy cluster Abell 780, shows a spectacular S-shaped morphology within the central 20~kpc. The symmetrical S-structure is also visible in the extended low frequency images at 74 and 330~MHz \citep{lane04}; the radio source extends approximately 340~kpc to the north and 190~kpc to the south. Modelling the entire source is computationally impractical and we have adopted the approach of modelling the innermost structures first in order to constrain jet parameters \citep{nawaz14a}, then utilising these parameters in models of the intermediate scale structure (this paper) and finally the large scale structure. It is noted here that, unless otherwise stated, all sizes and distances mentioned in this paper are deprojected distances assuming an approximate inclination angle of the jet axis of Hydra A to the line of sight of $\theta = 42^{\circ}$ which was estimated by \citet{taylor93} from the rotation measure asymmetry of Hydra A. The parameters used to estimate the source extent are provided in Table~\ref{t:dcal}.

\begin{table}
\caption{Parameters used to estimate the source extent.} 
\centering
\begin{tabular}{l c c}
 \hline
 Parameter & Value \\
 \hline 
 Redshift, $z$  &  0.054 \\
 Hubble constant, $H_0$   &  71                       \\ 
 Luminosity distance & 230~Mpc   \\
 kpc per arcsec  & 1.1 \\
 Angle between the jet and the line of sight & $42^{\circ}$ \\
 \hline 
\end{tabular}
\label{t:dcal}
\end{table}

In Paper I, we commenced our study of Hydra A focussing on the central 10~kpc structure of the northern jet. We studied the kinetic power of the Hydra A jets and two key features of the inner 10~kpc of the northern jet: i) the radius of the collimated jet as a function of distance from the core, and ii) two bright knots at approximately 3.7~kpc and 7.0~kpc. Since the jet is mildly curved within 10~kpc, we used two-dimensional axisymmetric simulations to model the inner two bright knots as biconical reconfinement shocks. By fitting the knot location and the radius profile of the observed jet with our models we estimated the jet velocity at 0.5~kpc to be approximately 0.8c, the jet over-pressure ratio  with respect to the ICM to be approximately 5, and the jet density parameter  $\chi_{\rm jet} = \rho_{\rm jet}/(\varepsilon_{\rm jet} + p_{\rm jet})\approx 13$, where $\rho_{\rm jet}$, $\varepsilon_{\rm jet}$ and $p_{\rm jet}$ are the rest mass density, energy density and the pressure of the jet, respectively.

\begin{figure}
\centering
\includegraphics[width=\linewidth]{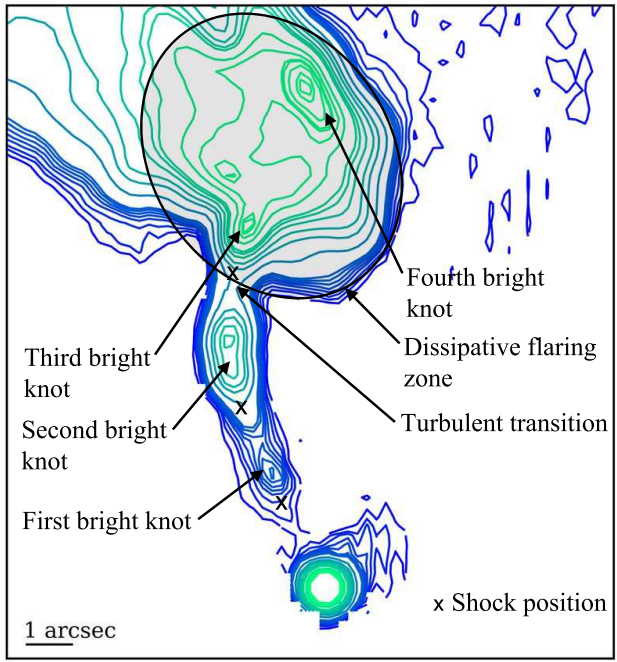}
\caption{Radio intensity map of the central 20~kpc of the Hydra A northern jet at 4.635 GHz. Contour levels are at 1.5, 2.7, 3.7, 5.1, 6.3, 7.5, 8.8, 10, 21, 37, 51, 72, 90, 103, 150, 180, 200, 205, 220, 240 and 249 mJy arcsec$^{-2}$. Four bright knots are marked with black arrows. The locations of the biconical reconfinement shocks which we interpret as the cause of the bright knots \citep{nawaz14a} are marked with $\times$. The turbulent transition of the jet starts near the third reconfinement shock. The turbulent flaring zone, shaded by an ellipse, is the beginning of a wide plume.}
\label{f:obs}
\end{figure}

In the present study we address the following additional key features of the inner 30~kpc of the northern Hydra A jet: i) the curved jet morphology, ii) two additional bright knots beyond 10 kpc and iii) the turbulent transition of the jet to a dissipative plume. In Fig.~\ref{f:obs} we show the 4.635~GHz radio structure of the northern jet and label these features. Detailed description of the 4.635~GHz VLA data are provided in \citet{taylor90}. In order to model the curved jet morphology we perform three dimensional relativistic hydrodynamical simulations of a precessing jet based on the jet and interstellar medium (ISM) parameters derived in Paper I. 

In outline, our model for the Hydra A northern jet is as follows: The initially conical, precessing jet expands through the ambient medium. The curvature of the jet is caused by its precessional motion.
As a result of the interaction between the jet and the ISM a series of biconical reconfinement shocks which manifest themselves as bright knots appear along the jet axis as in our straight jet models in Paper I. The initially supersonic jet is decelerated by the combined effect of reconfinement shocks (Nawaz et al. 2014, Perucho et al. 2007) and turbulent mixing of the jet boundary with the ambient medium \citep{perucho14a, deyoung93, bicknell84}; this process is enhanced by the instabilities resulting from the shear induced by the shocks. The shock deceleration is counterbalanced somewhat by the decrease in the atmospheric pressure along the direction of jet propagation. As a result, the jet is collimated by a series of reconfinement shocks until the turbulent jet boundary propagates significantly into the interior of the jet, causing it to flare near the third bright knot.
The turbulent jet hits the dense cocoon wall near the fourth knot and the backflowing jet plasma creates a strong turbulent dissipative zone (marked by a shaded ellipse in Fig.~\ref{f:obs}). Beyond this dissipative region, the jet develops into a wide buoyant plume.

The paper is structured as follows. In Section~\ref{s:model} we present a detailed description of our model of the precessing jet-ISM interaction. Sections~\ref{s:results} and Section~\ref{s:plume} present the results of our medium and high resolution models respectively. We summarise and discuss our results in Section~\ref{s:discussion}.

%
%
\section{Details of the model}\label{s:model}

The motivation for this study is to understand the dynamical interaction of the inner Hydra A northern jet with the interstellar medium and cluster environment and to understand the reason for the source morphology. Therefore, we mainly focus on the features of the inner 30~kpc including i) the curved jet ii) the four bright knots at approximately 3.7~kpc, 7.0~kpc, 11.0~kpc and 14.0~kpc from the core iii) the turbulent transition of the jet to a plume at approximately 10~kpc from the core, and iv) the bright radio emission region at approximately 10 to 20~kpc from the core. 

In Paper I, using axisymmetric straight jet simulations we modelled the first two bright knots of the northern jet as biconical reconfinement shocks. In this paper we develop this model by introducing jet precession; this necessitates three dimensional simulations.  According to our model, the jet is initially ballistic and conically expands in the first 0.5~kpc. It then starts to interact with the ISM and is collimated by the ambient pressure. A series of bright knots are produced along the jet trajectory at the locations of the biconical reconfinement shocks. 

The initially supersonic jet is decelerated significantly by the first two reconfinement shocks and the jet starts to form a turbulent plume at approximately 11~kpc from the core. The jet strongly interacts with the ISM and produce further reconfinement shocks at approximately 11~kpc and 16~kpc  from the core. Some jet plasma is deflected by the dense cocoon wall near the fourth knot and a highly turbulent zone is established in the region of approximately 11-20~kpc from the core. 

In this investigation of the effects of jet precession, our modelling strategy is as follows: We first conduct a medium resolution parameter space study with jet parameters (radius, kinetic power, velocity, density and pressure) derived from the best fit axisymmetric model of Paper I, a range of precession periods and two values of the precession cone angle. For the parameter space study we model the inner 20~kpc of the northern jet focussing on the curvature of the jet, the four bright knots and the turbulent transition of the jet. We construct synthetic surface brightness images of the models and compare the source morphology obtained from our models with the observations. Matching the key features of the inner 20~kpc of the northern jet, we select a best matching model and hence estimate the precession period and precession angle of the Hydra A northern jet. Using the estimated precession period and precession angle, we further model the inner 30~kpc of the northern jet at higher resolution and study the turbulent flaring zone, jet to lobe transition and formation of a wide plume. 

%
%
\subsection{Precession geometry, simulation setup and parameters}
\begin{figure}
\centering
\includegraphics[width=\linewidth]{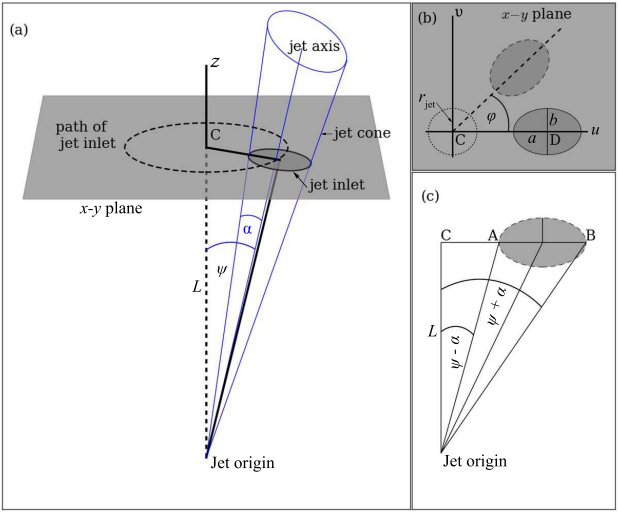}
\caption{Geometry of the precessing jet model. Panel (a) shows  the conical jet originating at a distance $L$ below the $x-y$~plane of the computational domain. The precessing jet cone intersects the $x-y$~plane in an elliptical jet inlet which moves on the (dashed) circular path. The coordinates $(u,v)$, defined by the intersection of the cone and the $x-y$~plane at a precession azimuth $\phi = 0^{\circ}$ and an arbitrary $\phi$ are shown in panel (b). The dotted circular line is the intersection of the cone when the precession angle $\psi = 0^{\circ}$.
The jet semi-minor axis of the jet inlet $b$ is equal to the jet radius $r_{\rm jet}$.  In panel (c) the angles defined by the lines joining the jet origin and the left and right edges of the inlet ellipse are shown. These define the semi-major axis of the ellipse.}
\label{f:mod}
\end{figure}
The geometrical configuration of our precessing jet model for the Hydra A northern jet is shown in Fig.~\ref{f:mod}.  The jet originates near the central black hole (marked as the jet origin in panel (a)) and is initially ballistic and conically expanding \citep{komissarov98, krause12, nawaz14a}. It precesses around the $z$-axis with a precession period $P$ and a precession angle $\psi$. The best fit axisymmetric model (presented in Paper I) gives a jet radius $r_{\rm jet} = 0.1$~kpc at $L=0.5$~kpc away from the black hole. The half angle of the jet cone is then $\alpha = \tan^{-1}(r_{\rm jet}/L) = 11.3^{\circ}$.

The jet cone intersects the $xy$~plane at a distance $L$ from the central black hole in an ellipse. As a result of precession the elliptical jet inlet follows a circular path (marked in panel (a)) on the $xy$~plane. The elliptical jet base is determined from the geometry shown in panels (b) and (c) of Fig.~\ref{f:mod} as described below.  

Let $(u,v)$ be a rotating frame fixed on the elliptical jet inlet. The semi-major axis $a$ and semi-minor axis $b$ of the ellipse lie on the $u$ and $v$ axes respectively (see panel (b) of Fig.~\ref{f:mod}). The centre of the ellipse lies at 
\begin{eqnarray}
u_0 = L[\tan(\psi+\alpha) + \tan(\psi - \alpha)]/2, \\
v_0 = 0.
\end{eqnarray}

From the geometry described in panels (b) and (c) we obtain 
\begin{eqnarray}
a &=& L[\tan(\psi + \alpha) - \tan(\psi -\alpha)]/2, \\
b &=& r_{\rm jet}.
\end{eqnarray}
Therefore, in the rotating $(u,v)$ coordinate system the jet inlet is defined by 
\begin{equation}
(u - u_0^2)/a^2 + v^2/ b^2 \leq 1.
\end{equation}

For a clockwise rotation of the jet inlet the coordinates $uv$ are related to the computational coordinates $xy$: 

\begin{eqnarray}
u &=& x\cos \phi - y \sin \phi, \\
v &= & x \sin \phi + y \cos \phi \,.
\end{eqnarray}

Here, $\phi = 2\pi t/P$ is the azimuthal angle of precession. Regarding the direction of precession, we refer to \citet{hamer14} who discovered a large rotating disk at the centre of the Hydra A cooling flow cluster. Assuming that the accretion disk rotates in the same direction as the outer gas disk and further assuming that the jet precession is associated with the precession of the inner accretion disk and black hole, we adopt a clockwise precessing jet in our model.

In order to avoid reverse shocks affecting ghost zones at the jet inlet, we initialise the jet in the computational domain with a semi-ellipsoidal cap above the jet inlet with semi-principal axes $a, \ b$ and $c (= a)$. 

The input jet parameters, chosen from the best-fit axisymmetric model presented in Paper I, are the jet kinetic power $P_{\rm jet} = 1\times 10^{45}$~erg s$^{-1}$, the jet over-pressure ratio $p_{\rm jet}/p_{\rm a} = 5$, the jet velocity $\beta = 0.8$, and the jet density parameter $\chi_{\rm jet} = 12.75$. We explore a range of values for the precession period $P = 1, 5, 10, 15, 20, 25$~Myr and the precession angle $\theta = 15^{\circ} \rm \ and \ 20^{\circ}$. The grid of models is presented in Table~\ref{t:mod}. Since the radiative cooling time of the thermal environment is large compared to the simulation time, we do not include cooling in our models. 

As described in Paper I, the three-dimensional cluster environment is constructed using analytical fits for the density, pressure and temperature data derived from the X-ray data presented by \citet{david01}.

We set up the computational grid as follows: For simulations employing a smaller grid 
($20 \times 20 \times20 \> \rm kpc$) and medium resolution we use a $156 \times 156 \times 156$ uniform grid for the inner 5~kpc, thereby obtaining six cells across a jet diameter, corresponding to approximately 28 cells per jet cross-section. For the remaining computational domain we use a stretched grid with 100 additional cells along each of the coordinate directions. In the simulations employing a larger volume ($30 \times 30 \times30 \> \rm kpc$) and high resolution we resolve the jet base by 12 cells ($\approx 113$ cells per jet cross section) using a $256 \times 256 \times 256$ uniform grid for the inner $5 \times 5\times 5 \> \rm kpc$ and a stretched grid with 100 additional cells along each of the coordinate directions.

The simulations were performed using the public domain relativistic hydrodynamic code PLUTO\footnote{http://plutocode.ph.unito.it} \citep{mignone07}, which solves the relativistic gas dynamical equations using a Godunov scheme implemented with a finite volume algorithm and a relativistic Riemann solver to calculate the fluxes. 

\begin{table*}
\centering
\caption{Grid of precessing jet-ICM interaction model. }
\begin{threeparttable}
\begin{tabular}{*{7}{c}}
\hline \hline

Model &   Period & Precession &  Grid size  \\
      & (Myr)  & angle (degrees) &  (kpc~$\times$~kpc$\times$~kpc)\\ 
 \\ \hline
   A & 1.0 & 20 & 20~$\times$~20~$\times$~20\\
   B & 1.0 & 15  & 20~$\times$~20~$\times$~20\\
   C & 5.0  & 20  & 20~$\times$~20~$\times$~20\\
   D & 10.0 & 20 & 20~$\times$~20~$\times$~20\\
   E & 15.0 & 20 & 20~$\times$~20~$\times$~20\\
   F & 20.0 & 20 & 20~$\times$~20~$\times$~20 \\
   G & 25.0 & 20 & 20~$\times$~20~$\times$~20\\
   H\_high\_res & 1.0 & 20 & 30~$\times$~30~$\times$~30 \\
\hline
\end{tabular}
\label{t:mod}
\end{threeparttable}
\end{table*}

%
%
\subsection{Synthetic surface brightness}\label{s:sb}
\begin{figure*}
\centering
\includegraphics[width=\textwidth]{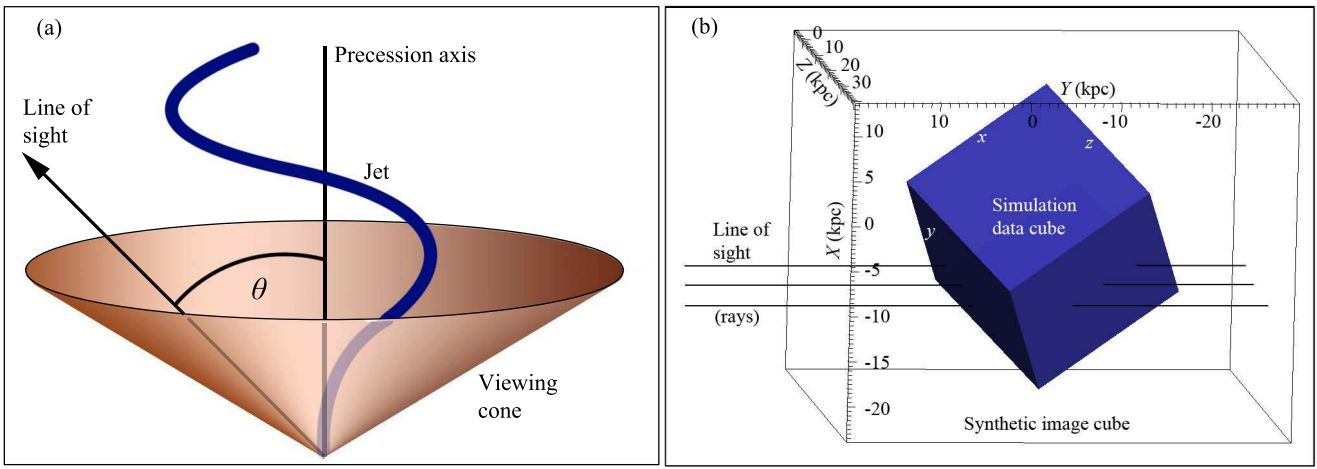}
\caption{Dependencies of the jet morphology on the line of sight and the viewing direction. (a) A cartoon of a spiral jet, an arbitrary line of sight and a viewing cone with cone axis aligned with the jet axis and cone angle equal to the line of sight angle are shown. Any line of sight lying on the viewing cone has the same inclination $\theta$ with the jet axis. The observed source morphology depends on both the line of sight inclination $\theta$ and the viewing direction. (b) The image cube, the data cube and the line of sight (marked by rays) are shown. The data cube is rotated with respect to the image cube to obtain any line of sight and a viewing direction.}
\label{f:con}
\end{figure*}

In order to compare the morphologies derived from the models with the radio observations we produce synthetic surface brightness images for each model. Following \citet{sutherland07}, we use a synchrotron rest-frame emissivity $j_\nu \propto p^{(3+\alpha)/2}$ where  $\alpha$ is the spectral index (Flux density $\propto \nu^{-\alpha}$).  In this formulation, the magnetic pressure is assumed to be proportional to the total particle pressure. The northern Hydra A jet approaches the observer; hence the emissivity $j_\nu$ is modified by the Doppler factor $\delta = 1/\Gamma(1 - \beta \cos\theta]$, where $\Gamma$ is the bulk Lorentz factor and $\theta$ is the angle between the jet axis and the line of sight. 
 
In addition, in order to isolate the jet plasma from the ambient medium we use a tracer $\lambda$, which is the mass fraction of jet plasma at each cell. We initialise the jet plasma with a value $\lambda = 1$. Hence the emissivity $j_\nu$ becomes
\begin{equation}
j_\nu = \lambda \delta^{2+\alpha}p^{(3+\alpha)/2}
\end{equation}
Integrating the synchrotron emissivity along rays, parallel to the line of sight, $I_\nu = \int j_\nu \> ds$, we obtain images of the synthetic surface brightness of the modelled jets. 

We note that the observed source morphology depends on both the angle between the jet axis and the line of sight and the viewing direction in azimuth. For instance, Fig.~\ref{f:con} shows an arbitrary spiral jet structure about the jet axis and an arbitrary line of sight (making an angle $\theta$ with the jet axis). In this figure a viewing cone is also shown. The axis of the viewing cone lies along the jet axis and its cone angle is equal to the inclination of the line of sight $\theta$. Any line of sight lying on the viewing cone has the same inclination $\theta$ but different azimuthal direction. It is clear from this figure that the jet morphology is different if either $\theta$ or the azimuth direction, or both, change. Therefore, we scan the synthetic images for different lines of sight and azimuth until we obtain the best match of the synthetic surface brightness to the observations.

In using the VisIt visualisation software\footnote{https://wci.llnl.gov/simulation/computer-codes/visit/}, it proved to be expedient to work with a fixed image cube and to rotate the computed emissivity cube in order to investigate the dependence of the synthetic image on viewing direction. The data cube is rotated so that the line of sight along which the surface brightness is calculated is the $Y$-axis of the image cube. We perform four successive rotations of the data cube ($xyz$) with respect to the image cube ($XYZ$) to obtain a desired line of sight and viewing direction. First two rotations of the data cube, $\phi$ (azimuth angle of the jet) and $\psi$ (half precession cone angle), bring the jet axis along the line of sight $Y$-axis. The following two rotations of the cube, $\chi$ (rotation about the jet axis; $0 < \chi < 2 \pi$) and $\theta$ (the angle between the jet and the line of sight) set the viewing direction. Details of the transformations are presented in Appendix~\ref{A:trans}. 

Let $\textbf{v}'$ and $\textbf{v}$ be the velocity vector of the fluid in the image cube and data cube respectively. Then the velocity $\textbf{v}'$ is given by 

\begin{equation}
\textbf{v}' = \textbf{R} \textbf{v}
\end{equation}
where $\textbf{R}$ is the  matrix of the transformation from the data cube to the image cube (see Appendix~\ref{A:trans} for the description of $\textbf{R}$).

The angle between the line-of-sight ($Y$-axis) and the fluid velocity at a cell is given by 
\begin{equation}
\theta' = \cos^{-1}v'_Y / v'
\end{equation}
where $v'_Y$ and $v'$ are the $Y$ component and magnitude of the velocity in the image cube, respectively. 

To obtain the correct Doppler factor for each cell we use the angle $\theta'$ when determining the Doppler beaming factor $\delta = \Gamma^{-1} (1 - \beta \cos \theta^\prime)^{-1}$.

Since we are considering the Doppler beaming for individual cells in the simulation data cube, changing the line of sight or viewing direction not only changes the radio morphology of the synthetic image, but the relative brightness of different regions in the source as well. In \S~\ref{s:plume} we discuss the change in the apparent morphology of the simulated jet resulting from the change in the viewing direction.

%
%

\section{Simulation Results}\label{s:results}

In this section we present the results of our three dimensional simulations.
First, we present our medium resolution parameter space studies and discuss the difference in the jet morphologies for different precession periods and precession angles. Matching the simulated and observed jet-lobe morphologies of the inner 20~kpc, in particular, the curvature of the jet and the four bright knots and their locations, we select a best matching model and hence estimate the precession period and precession angle for the Hydra A northern jet. We then estimate the Mach number of the advancing forward shock and discuss the heating of the ISM by the AGN. 

\subsection{Jet curvature}
\label{curvature}
\begin{figure*}
\centering
\includegraphics[width=\textwidth]{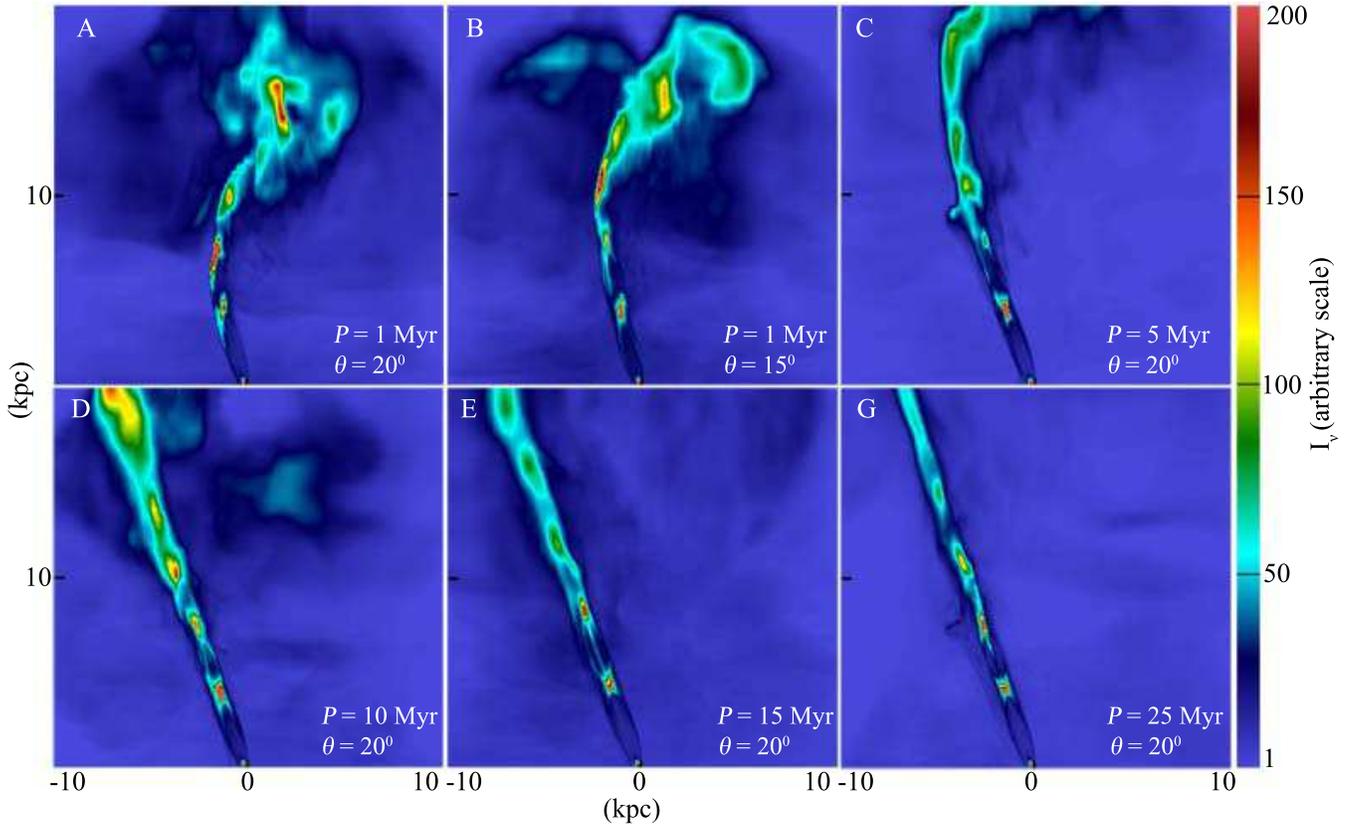}
\caption{Synthetic surface brightness of models A, B, C, D, E and G. The snapshots are chosen for a simulation time at which the jet is fully developed in the computation domain. The line of sight angle for each panel is 90$^{\circ}$ and the viewing directions are chosen in such a way that the jets show maximum curvature. }
\label{f:cur}
\end{figure*}

The most useful discriminant between the models is the jet curvature. Fig.~\ref{f:cur} shows the synthetic surface brightness images for models A, B, C, D, E, and G. The snapshots are taken when the jet is fully developed in the computational domain. Notwithstanding the dependence of morphology on viewing angle and the location of the jet in its precession cone, these snapshots are very informative in discriminating between different precession periods and angles.

In Fig.~\ref{f:cur} it is evident that the curvature of the jet depends strongly on the precession period and increases as the precession period decreases. Models with longer precession periods produce straight jets within the first 10~kpc. For example, jets produced by the models C, D, E, and G with precession periods 5, 10, 15 and 25~Myr, respectively, are straight in the inner 10~kpc. The jet with a precession period of 1~Myr and a precession angle of 15$^{\circ}$ is also nearly straight within this region, only showing a bend beyond approximately 10 kpc. We see a mild curvature inside 10~kpc for model A, which has a precession period of 1~Myr and a precession angle of 20$^{\circ}$. This curvature is comparable to the curvature of the Hydra A northern jet. Therefore, on the basis of this curvature comparison alone, model A is our best match for Hydra A. This choice is confirmed by other observational features of the inner 20~kpc of the the northern jet reproduced by this model, such as the correct number of knots and their locations as well as the turbulent transition of the jet to a plume. In the following we discuss the morphological features developed in model A and compare them with the inner 20~kpc Hydra A northern jet morphology.

\subsection{Bright knots and the turbulent transition of the jet}
\begin{figure*}
\centering
\includegraphics[width=\textwidth]{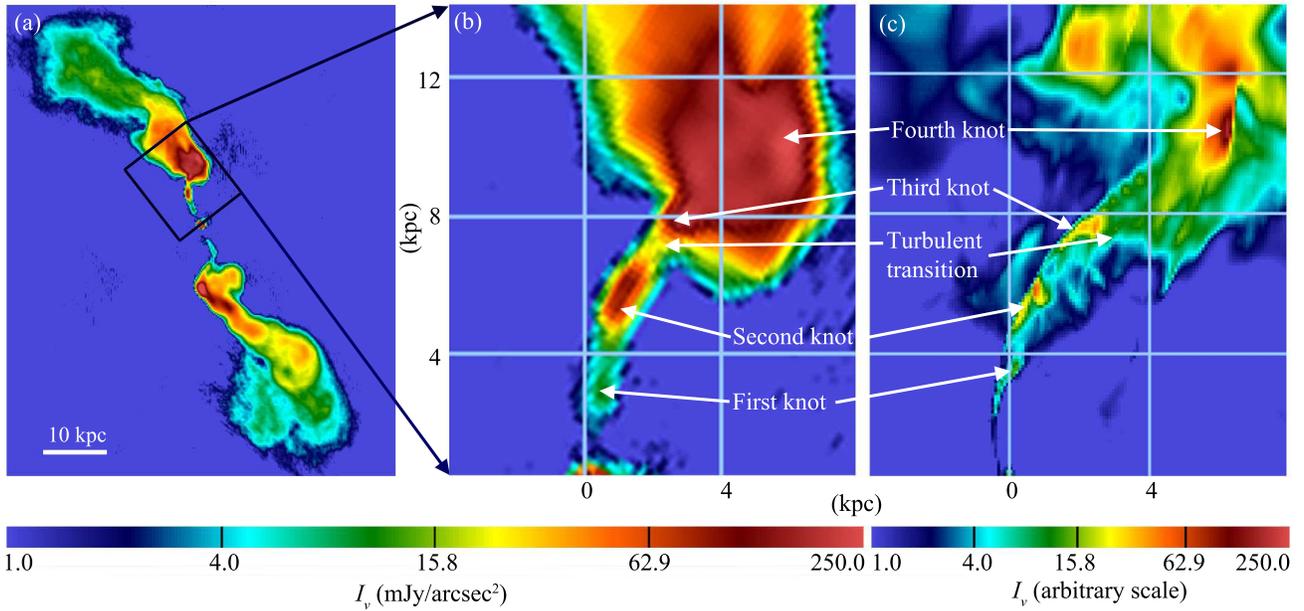}
\caption{ A comparison between the source morphology of the best match model and the observational data by \citet{taylor90}. Panel (a): Total intensity image of Hydra A at 4635~Mhz reproduced from the original data presented in \citet{taylor90}. Panel (b): Zoom-in of the section of the northern jet marked by a rectangle in the left panel. Panel (c): Synthetic total intensity image of the simulated jet at optimal viewing parameters, the line-of-sight angle, $\theta = 42^{\circ}$ and the azimuthal angle $\chi = 45^{\circ}$ (see Appendix~\ref{A:trans} for the definition of $\theta$ and $\chi$).  }
\label{f:hyd}
\end{figure*}

\begin{figure*}
\centering
\includegraphics[width=\textwidth]{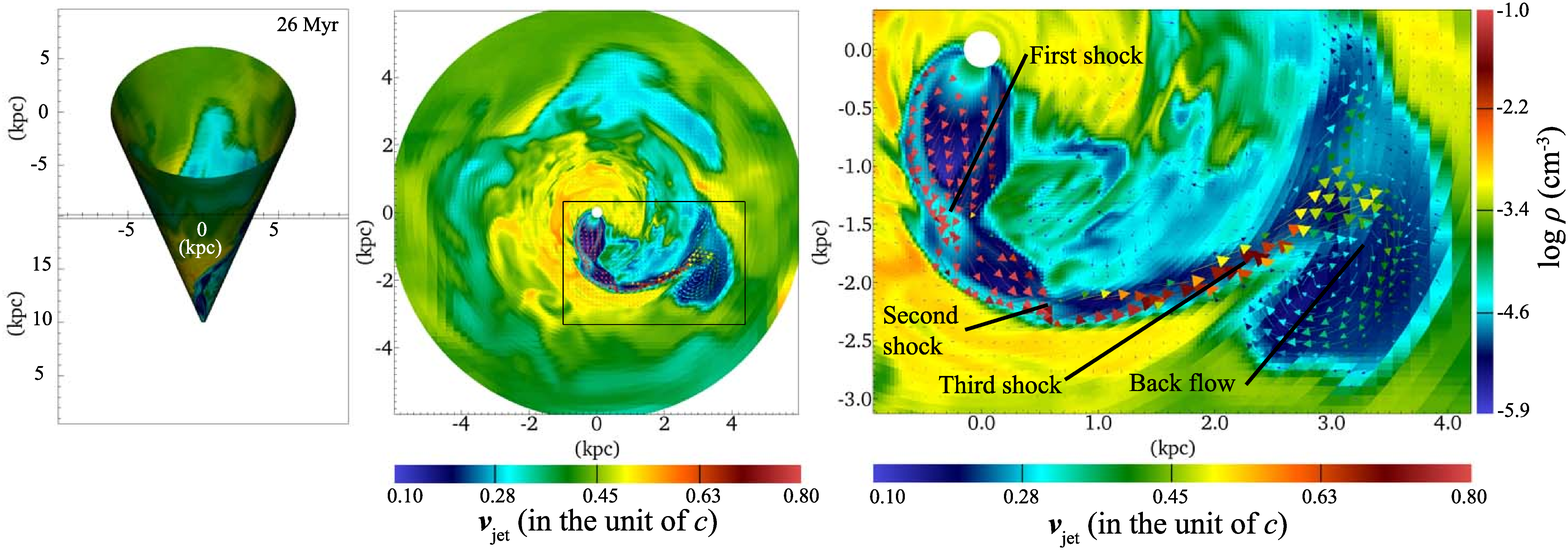}
\caption{ Conic slice (cone angle 17$^{\circ}$ and cone axis aligned with the $z$-axis) of the logarithmic density image overlaid with the flow velocity vectors of the best-matching model (model A) at a simulation time 26 Myr (left panel). The middle panel shows the projection of the cone onto the $x-y$ plane. The right panel is a zoom in image of the region marked by a rectangle in the middle panel.  }
\label{f:fdir}
\end{figure*}

Fig.~\ref{f:hyd} compares the simulated jet of model A (panel (c)) and the inner 14~kpc (projected) of the Hydra A northern jet (panel (b)). Panel (a) of this figure shows the total intensity image of Hydra A at 4635~Mhz~reproduced from the original data presented in \citet{taylor90}. Panel (b) is a zoom-in of the section of the northern jet marked by a rectangle in panel (a).  Panel (c) shows the simulated jet with optimal viewing parameters for Hydra A. The viewing parameters for panel (c) are as follows: the line-of-sight angle, $\theta = 42^{\circ}$ and the azimuthal angle, 
$\chi = 45^{\circ}$ (see Appendix~\ref{A:trans} for the definition of $\chi$). We choose $\theta = 42^{\circ}$ as the line of sight angle for consistency with Paper~1 and the approximate value derived by \citet{taylor93}. The viewing direction (angle $\chi$) is determined by trial and error and we found $45^{\circ}$ to be an optimal viewing direction. The length scales in each panel are projected distances from the core.

The moderately over-pressured precessing jet interacts with the ambient medium and produces three reconfinement shocks at approximately 2.1~kpc, 4.5~kpc, 7.2~kpc from the core. Since the synchrotron emissivity $j_{\nu} \propto p^{(3+\alpha)/2}$, downstream of the reconfinement shocks the pressure and therefore the surface brightness increase producing three bright knots (marked by arrows in panel (c)). A fourth knot is produced at approximately 9.2~kpc from the core where the jet hits the cavity wall. We see that the locations of the bright knots produced with this model agree well with the locations of bright knots in the Hydra A northern jet located at approximately 2.3~kpc, 4.7~kpc, 7.4~kpc and 8.7~kpc (projected) from the core (marked by arrows in panel (b)). In the simulated jet a turbulent transition occurs approximately near the third bright knot (marked in panel (c)), which is also consistent with the observations (marked in panel (b)). Later, in the larger scale model (run $H\_\rm{high}\_\rm{res}$) we see that this turbulent transition creates a wide plume similar to the plume of the Hydra A northern jet.

\subsection{The turbulent flaring zone}
\label{flaring}
Fig.~\ref{f:fdir} shows the logarithmic density of run A (at a simulation time 26 Myr) sliced by a cone with a cone angle of 17$^{\circ}$ (left panel) and cone axis aligned with the precession axis ($z$ axis). To obtain a clear view of the jet and the flow direction the cone is projected onto the $x-y$~plane (right panel of Fig.~\ref{f:fdir}) and overlaid with the flow velocity vectors. A zoomed-in image of the region marked by a rectangle in the middle panel is shown in the right panel. We note here that, although the precession angle in model A is 20$^{\circ}$, the jet is mostly visible along the conic slice with a cone angle 17$^{\circ}$. This is the result of the reflective boundary condition at the lower $z$ boundary. The reflection of the back flow on the side of the jet closest to the boundary pushes the jet towards the precession axis. Therefore, the jet is maximally visible along a conic slice with cone angle less than $20^{\circ}$. 

In Fig.~\ref{f:fdir} we see that after the turbulent transition of the jet some jet plasma hits the dense cocoon plasma and produces a strong back flow (shown in the right panel). This turbulent back flow  establishes a flaring region. Such a flaring region is apparent in the observed source at approximately 8 to 14~kpc (projected) from the core in the northern jet of Hydra A.
 Moreover, in the polarisation image of Hydra A \citep{taylor90} we see that the polarisation  falls from 40$\%$ (in the collimated jet) to 10$\%$ in the flaring region. This reduction in polarisation suggests that the flaring region of the northern jet is turbulent. This is consistent with our simulations.  
 
 From the above discussion we see that model A can produce the correct curvature of the jet, four bright knots, the turbulent transition of the jet and the turbulent flaring zone of the inner 20~kpc of the Hydra A northern jet. Therefore, our best match model is run A and hence our estimation for the precession period and precession angle are approximately 1~Myr and 20$^{\circ}$, respectively, for the jets of Hydra A. We provide further justification for these estimates by modeling the inner 30~kpc of the northern jet, focussing on its its jet-plume morphology. In \S~\ref{s:plume} we present the high resolution 30~kpc scale model. 
 
%
%
\subsection{Forward shock} 
\begin{figure*}
\centering
\includegraphics[width=\textwidth]{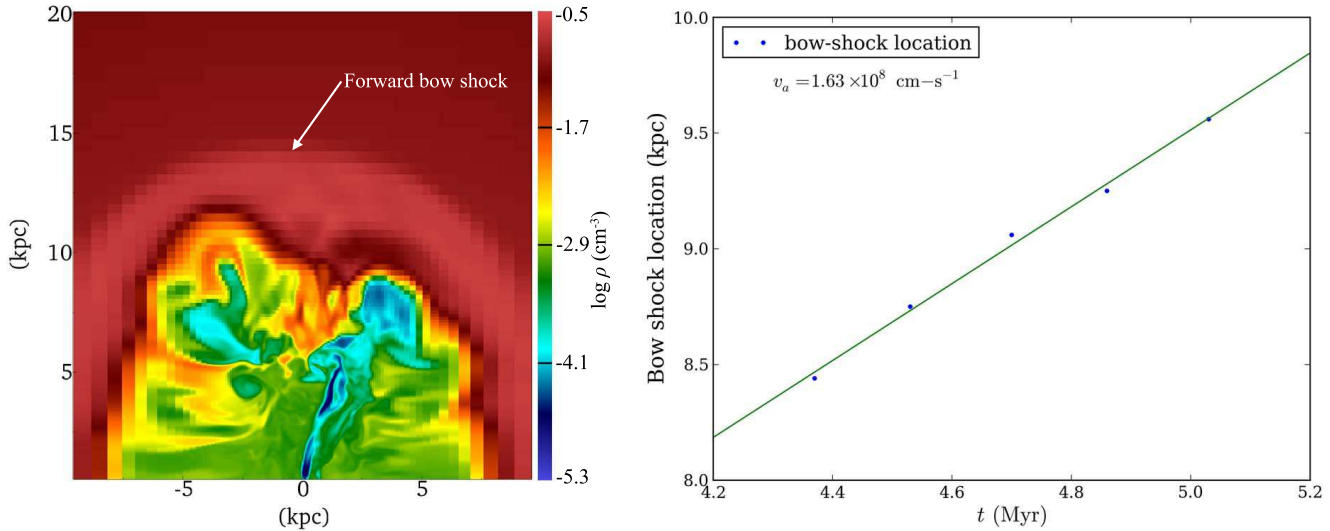}
\caption{ Left: Midplane slice of the logarithmic density snapshot of model A. The forward bow shock is marked by an arrow in this panel. Right: Locations of the forward shock at five different time steps (points) . A least square linear fit (line) gives a shock advance speed $\approx$ 1630 km s$^{-1}$. }
\label{f:fsh}
\end{figure*}
In our best-matching model, the radio jet-ICM interactions are bounded by an advancing forward shock. Here we estimate its Mach number. 

The forward bow shock is shown in the logarithmic density snapshot of model A (left panel of Fig.~\ref{f:fsh}). In the right panel of Fig.~\ref{f:fsh} we trace the location of the apex of this shock along the $z$-axis at five different time steps. Fitting a least square line to the shock locations we obtain a shock advance speed $\approx  1630$ km s$^{-1}$ of the forward shock. The sound speed at appleroximately 15~kpc from the core is $\approx 880$ km s$^{-1}$. Hence, the Mach number of the forward shock is $\approx 1.85$. 
There is a mild pressure jump $\approx 3.4$ at the forward shock. The low Mach number and mild pressure jump indicate that the heating of the atmosphere by the radio AGN in its earlier stage is gentle. This is a required feature of the heating of cooling flows \citep{mcnamara12}. 

%

%
%

\section{Results from the high resolution larger volume model}\label{s:plume}
Here we present the results of our high resolution larger volume $30 \times 30 \times 30 \> \rm kpc$ model (run H$\_{\rm high}\_{\rm res}$) and discuss the source morphology in more detail. First we show that this model can again successfully reproduce the key features of the inner 30~kpc of the Hydra A northern jet. Then we discuss the strong dependency of the source morphology on the azimuthal angle $\chi$.

\begin{figure*}
\centering
\includegraphics[width=\textwidth]{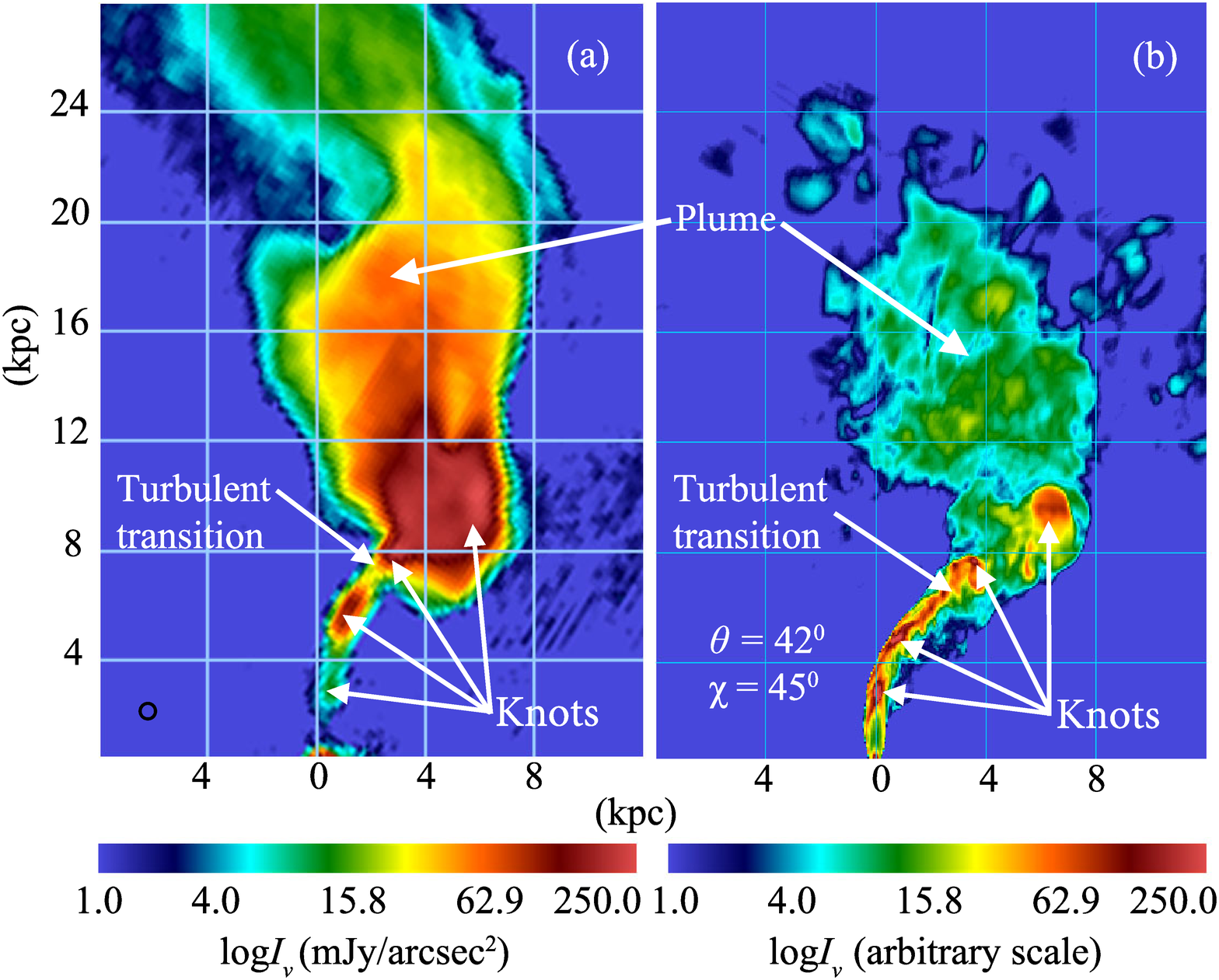}
\caption{A comparison between the jet-lobe morphology of the Hydra A northern jet and the best matching simulated jet. Panel (a): Total intensity image of Hydra A at 4635~Mhz with $0''.6$~resolution, reproduced from the original data presented in \citet{taylor90}. The beam size is shown in the lower left corner. Panel (b): The synthetic total intensity image of the simulated jets (run H$\_ \rm high \_ res$) with directional parameters $\theta =  42^{\circ}$ and $\chi = 45^{\circ}$. The length scales shown are projected distances from the core.} 
\label{f:plu}
\end{figure*}

\subsection{Development of the plume}
Fig.~\ref{f:plu} compares the morphology of the simulated jet-lobe structure and the observed Hydra A northern jet (up to 30~kpc from the core). In panel (a) we have the total intensity image of the Hydra A northern jet at 4635~Mhz with $0''.6$~resolution (reproduced from the original data presented in \citet{taylor90}). Panel (b) shows the simulated source with the line-of-sight angle, $\theta =  42^{\circ}$, and the azimuthal angle, $\chi = 45^{\circ}$. In the simulated jet we see four bright knots along the jet flow. The first three bright knots are produced due to shock deceleration of the jet plasma by reconfinement shocks. A fourth bright knot is produced when the supersonic jet hits the cavity wall and sharply changes direction. The bright region between the second and the third knot is due to the effect of Doppler boosting of the shocked plasma at the jet boundary in this region. This bright region disappears if we look at the source from a different direction (see different panels of Fig.~\ref{f:chi}). The locations of the bright knots are approximately at 2.6, 4.6, 7.6 and 8.8~kpc from the jet base. These knot locations closely match the observed knot locations at approximately  2.3, 4.7, 7.4 and 8.7~kpc (projected distance from the core) shown by arrows in panel (a).

The initially collimated jet starts to widen after the second bright knot at approximately 7~kpc from the jet base. The jet completely flares to form a plume near the third bright knot. This is consistent with the jet flaring in the Hydra A northern jet at approximately 7.3~kpc and the formation of the plume.
We note that the flaring location is quasi-steady and sometimes shifts slightly due to the strong interaction of the jet and the ambient medium. However it is reestablished quickly at approximately 7 kpc (projected) from the core.

The precessing jet twists along the surface of the precession cone and hence produces a spiral jet-lobe morphology. However, the source morphology obtained from the surface brightness image of the simulated jet depends strongly on the viewing parameters, the line of sight $\theta$ and the viewing angle $\chi$. For a set of viewing parameters, $\theta = 42^{\circ}$ and $\chi = 45^{\circ}$ we obtain a Hydra A like morphology.

In the Hydra A northern jet, the flaring region within approximately 8 to 14~kpc (projected) where the plume starts, is bright compared to the inner collimated jet. The corresponding region in our model does not reach the same level of brightness. The flaring region is strongly turbulent (see \S~\ref{flaring}), and the amplification of the magnetic field resulting from this turbulence and the 
associated re-acceleration of electrons may be responsible for the increase of the source brightness. Since, our model is purely hydrodynamic, the amplification of the magnetic field is not reflected in the synthetic brightness images. In order to produce more accurate synthetic brightness images magnetohydrodynamic (MHD) models are required. 

We note here that the source morphology within 20~kpc in the larger $30 \times 30 \times 30 \> \rm kpc$ model, for example the jet curvature, the locations of the the knots and the position of the turbulent transition of the jet, is similar to that of the source morphology obtained in the smaller  model (run A). This implies that our simulations are converging within the inner 20 kpc volume and that the downstream evolution of a jet does not effect the upstream morphology. 

\subsection{Complex source morphology}
\begin{figure*}
\centering
\includegraphics[width=\textwidth]{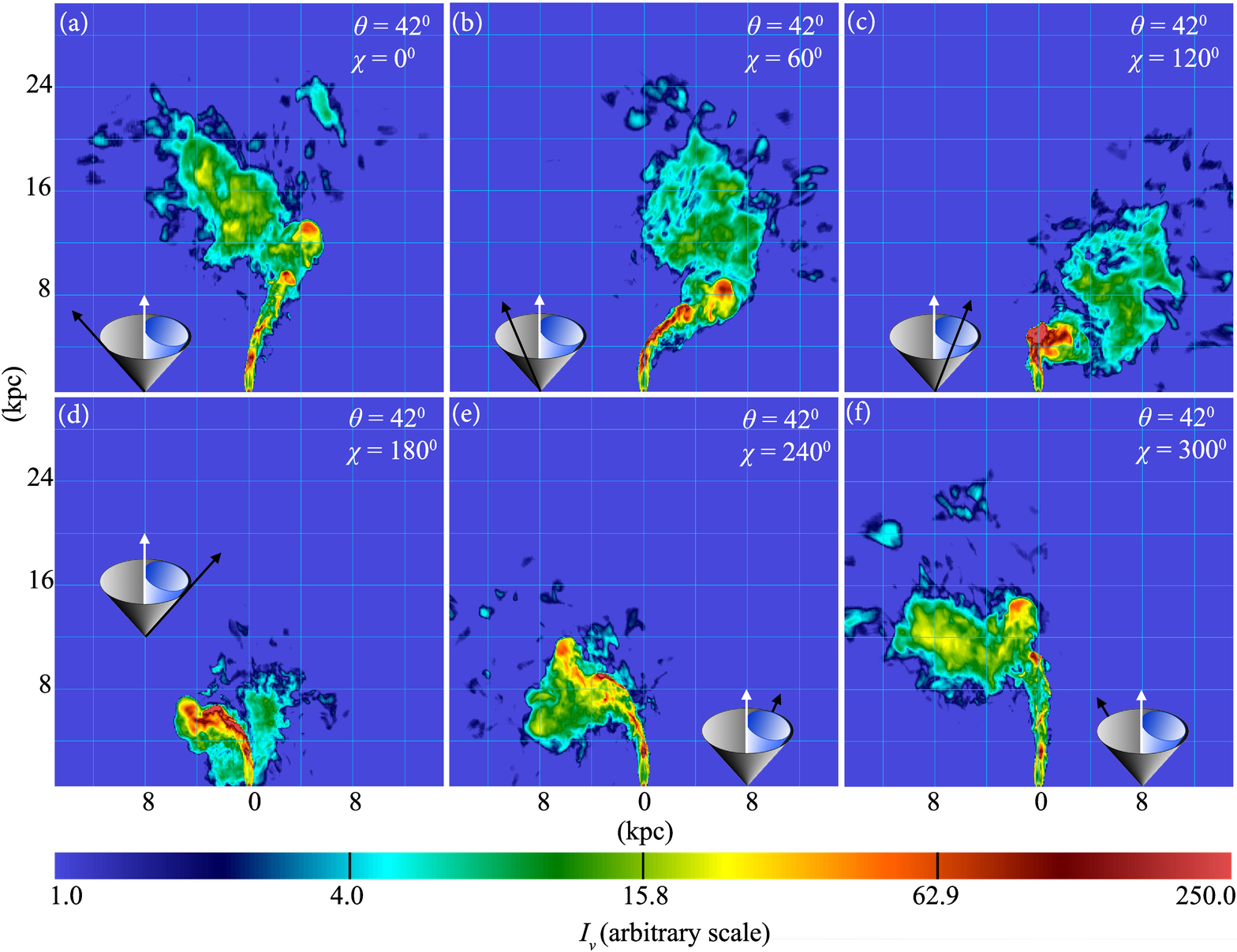}
\caption{Complex source morphologies depending on the viewing direction $\chi$. The line-of-sight angle for each panel is the same $\theta  = 42^{\circ}$, but the viewing direction $\chi$ varies from $0^{\circ}$ to $330^{\circ}$. For better understanding of the viewing parameters we put the viewing cone (black) and the precession cone (blue), with line of sight direction (black arrow) and the 
jet axis (white arrow) in each panel. }
\label{f:chi}
\end{figure*}

In a source with straight jets, the jet plasma propagates in an approximately cylindrically symmetric manner and the source morphology is independent of the viewing direction $\chi$ (see definition of $\chi$ in Appendix~\ref{A:proj}). However, as we discussed in \S~\ref{s:sb}, in the case of a precessing jet, the jet  has a spiral structure and the source morphology strongly depends on the viewing direction. Moreover, 
as a result of relativistic beaming,
the relative brightness between different regions of the jet and the lobe also depends on $\chi$.
Here we discuss the change in source radio  morphology when changing viewing parameters. The motivation for this is to show that the complex source morphologies, for instance S- X- or Z-symmetric structures of extragalactic radio jets may be attributed to jet precession.

Fig.~\ref{f:chi} shows the simulated source radio morphologies of the jet in run H$\_{\rm high}\_{\rm res}$ at a fixed line-of-sight angle $\theta =42^{\circ}$ but six different azimuthal angles $\chi = 0^{\circ}, 60^{\circ}, 120^{\circ}, 180^{\circ}, 240^{\circ} \rm and \ 300^{\circ}$. To help visualize the viewing parameters we have drawn a viewing cone (black) and a jet cone (blue), with the line of sight direction (black arrow) and the jet axis (white arrow) in each panel. The axes in each panel measure projected distances from the jet origin. In this figure we see that the source morphology varies greatly depending on the viewing direction. Certainly, many more complex source morphologies are possible for different lines of sights and viewing directions but we restrict ourselves to the case $\theta = 42^{\circ}$. 

In panel (a) the azimuthal angle is $\chi = 0^{\circ}$. At this angle, the line of sight is at the furthest possible distance from the precession cone (see the relative positions of the line of sight on the viewing cone and the jet cone). Hence, the extent of the radio structure is a maximum at $\chi=0$. A relatively bright lobe for these viewing parameters indicates that the lobe has a significant velocity component along this line of sight. This sort of jet-lobe structure is a feature of Z-shaped sources.

As $\chi$ increases, the line of sight gradually moves closer to the jet cone and hence the source morphology becomes more and more contracted and distorted (see panels (b) and (c)). The source morphology changes from a Z-symmetric structure to a S-symmetric one (see panel (b)). At $\chi = 180^{\circ}$ (panel (d)) the line of sight is closest to the precession cone. The line of sight intersects the lobe and the jet. This is nearly a view of the source along the jet axis and the spiral jet structure due to precession is seen clearly here. From viewing directions near 180$^{\circ}$ the source morphology resembles an X-shaped source. Further incrementing $\chi$, the line of sight moves away from the jet cone and the source morphology again turns towards Z-symmetric shapes.  

\section{Summary and Discussion}\label{s:discussion}
In Paper I, we modelled the inner two bright knots of the Hydra A northern jet with axisymmetric straight jet simulations. In this paper we have built on that model by incorporating jet precession and studying the three dimensional interaction of the jet with the intracluster medium. Our three-dimensional precessing jet model successfully reproduces the prominent features of the complex inner 30~kpc jet-lobe morphology on the northern side of Hydra A.  

We initially performed a parameter space study of precessing jet models, probing a range of precession periods (1, 5, 10, 15, 20 and 25~Myr) and two precession angles (15$^{\circ}$ and 20$^{\circ}$), while keeping the values of all other parameters fixed at the best-fit values found in Paper I.  We find that model A with a precession period of 1~Myr and a precession angle of 20$^{\circ}$ produces the correct jet curvature, the correct number of knots, and the jet to plume transition, all at approximately the correct locations. Therefore we select this model as our best-matching model for the Hydra A northern jet. Adopting the estimated precession period, precession angle and other jet parameters we enlarged the size of the computational domain from 20~kpc to 30~kpc, in order to better understand the complex jet-plume morphology. 

Our 30~kpc model successfully reproduces the following key features of the Hydra A northern jet: 
\begin{enumerate}
\item The correct curvature of the inner 7~kpc (projected) jet. 
\item Four bright knots along the propagation  of the jet. The first three bright knots appear behind biconical reconfinement shocks associated with the collimation of the jet by pressure of the ambient medium. A fourth bright knot appears at the location where the supersonic jet hits the side of the plume. The locations of the knots at approximately 2.1, 4.5, 7.2 and 9.3~kpc (projected) coincide well with the observed bright knots at approximately 2.3, 4.7, 7.4 and 8.7~kpc (projected). 
\item The turbulent transition of the jet to a plume at approximately $7$~kpc (projected); the observed transition location is at 7.5~kpc. The initially supersonic jet is significantly decelerated by the first two reconfinement shocks and the transition to turbulence begins after the second knot.  
\item The correct jet-lobe morphology. The structure of the jet and associated plume is a good morphological match to Hydra A.
\end{enumerate}

We have shown that the apparent radio morphology of a jet-lobe source strongly depends on the viewing direction $\chi$. For a particular source morphology we obtained Z, S or X-symmetric structures depending on the parameter $\chi$. Conversely, this result reinforces the idea that the radio sources exhibiting Z, S or X-symmetries have precessing jets. 

The larger $30 \times 30 \times 30 \> \rm kpc$
model produces a similar source morphology until 20~kpc from the jet origin to that obtained from the smaller $20 \times 20 \times 20 \> \rm kpc$ model.
This implies that the downstream evolution of the simulated jet does not effect the upstream structure. This result supports the bottom up approach  adopted in these two papers to study a very extended source like Hydra A, starting from the structures near the core, e.g. jet knots and jet curvature, to medium scale structures like the development  plume, and all the way to large scale structures like X-ray cavities and the outer bow shock. 

It is interesting to compare the observed and best fit model's projected angle $\omega$ on the sky between the jet axis and the precession axis (see Appendix~\ref{A:proj} for the definition and derivation of $\omega$). The line of sight angles $\theta = 42^{\circ}$ and $\chi = 45^{\circ}$ for the optimal view of the simulated jet gives a projected angle $\omega = 12.5^{\circ}$ between the jet axis and the precession axis on the sky plane. From the position angle of the VLBI northern jet, which is $\sim23^{\circ}$ \citep{taylor96} and the position angle of the kpc-scale molecular disk at the centre of the Hydra A galaxy, which is $\sim-74.5^{\circ}$ \citep{hamer14} we obtain a projected angle between the disk axis and the jet axis of $\sim7.5^{\circ}$. However, using emission lines of different molecules \citet{hamer14} estimated a range of position angles of the disk (from $-66^{\circ}$ to $-80^{\circ}$), which gives a
range for $\omega$ of $1^{\circ} - 13^{\circ}$. Hence, the value of $\omega$ we obtained from our model is similar to the upper limit of the observed value. We note here that the origin of the disk gas is most likely cooled material from the ICM, and the angular momentum of the disk should, therefore, reflect the angular momentum of the ICM gas that cooled and accreted toward the galaxy, implying that the atmosphere of Hydra A is far from static. Indeed, asymmetric bulk flows are expected from gas merger-induced ``sloshing'' \citep{zuhone10} or, as studied here, a precessing jet stirring the cluster atmosphere.

We estimate the Mach number of the forward bow shock to be $\approx 1.85$ from our 
optimal model.
This low Mach number shock and the associated weak pressure jump ($\approx 3.4$) suggest a gentle
and temporally extended
heating of the of the ICM by the radio AGN in its initial phases of evolution. The gentle heating of the ICM by the jet is consistent with the modern assessment of the heating of cooling flow clusters by AGN jets \citep{mcnamara12}. 

Inclusion of magnetic fields in this study would be interesting, mainly for the production of more realistic synthetic surface brightness images. For instance, magnetic field amplification in the turbulent flaring region (8-14~kpc) of the northern jet may be a possible explanation for the increase in brightness there - a purely hydrodynamic model does not capture this effect. We suspect that this is why, in our model, the brightness ratio between the initial jet (up to 8~kpc from the core) and the turbulent plume (8-14~kpc from the core) is not reproduced correctly. 

With regard to the southern jet of Hydra A, the initial 5~kpc of the trajectory of the jet is not well determined observationally. Therefore, modelling the southern side of the source as we have done for the northern side requires deeper high-resolution observations.

What is causing the jets to precess on a timescale as short as 1~Myr? The precession of the jets in Hydra A indicates that either the black hole or the inner disk, or both, are precessing with such a period. Much of the theoretical discussion of precessing disks has centred on the Bardeen--Petterson effect \citep{bardeen75a} wherein the combination of Lense-Thirring precession and accretion disk viscosity causes the disk to align with the angular momentum of the black hole and also to precess. The alignment and precession times are comparable \citep{scheuer96a,king05b} so that jet precession for several periods is an unlikely result of the Bardeen-Peterson effect. (In our simulations, the apparent morphology of the jet becomes comparable to the inner 30 kpc of the Hydra A northern source after 35 precession cycles of the jet with a steady precession period 1 Myr.) A more promising approach based on radiative warping \citep{pringle96a} or disk tearing, as suggested by \citet{nixon13b}, may be feasible. However, estimates of jet precession rates in AGN based upon either of these mechanisms is at an early stage, so that we defer consideration to a future paper.

\section*{Acknowledgments}

This research was supported by the Australian Research Council Discovery Project, DP140103341 \emph{The key role of black holes in galaxy formation}.  The computations were undertaken on the National Computational Infrastructure supercomputer located at the Australian National University. The research has made extensive use of the VisIt visualization and analysis tool VisIt (https://wci.llnl.gov/codes/visit). We thank Professor Gregory Taylor for providing us with the radio data of Hydra A used for Fig.~\ref{f:obs}, Fig.~\ref{f:hyd} and Fig.~\ref{f:plu}. We are grateful to the referee for a careful reading of the paper and for making useful suggestions for improvements.

\bibliographystyle{mn2e}
\bibliography{mohammad,gvbrefs}
%
%
\appendix
\section{Transformations associated with the rotations of points of the simulation data cube with respect to the synthetic image cube}\label{A:trans}
\begin{figure*}
\centering
\includegraphics[width=\textwidth]{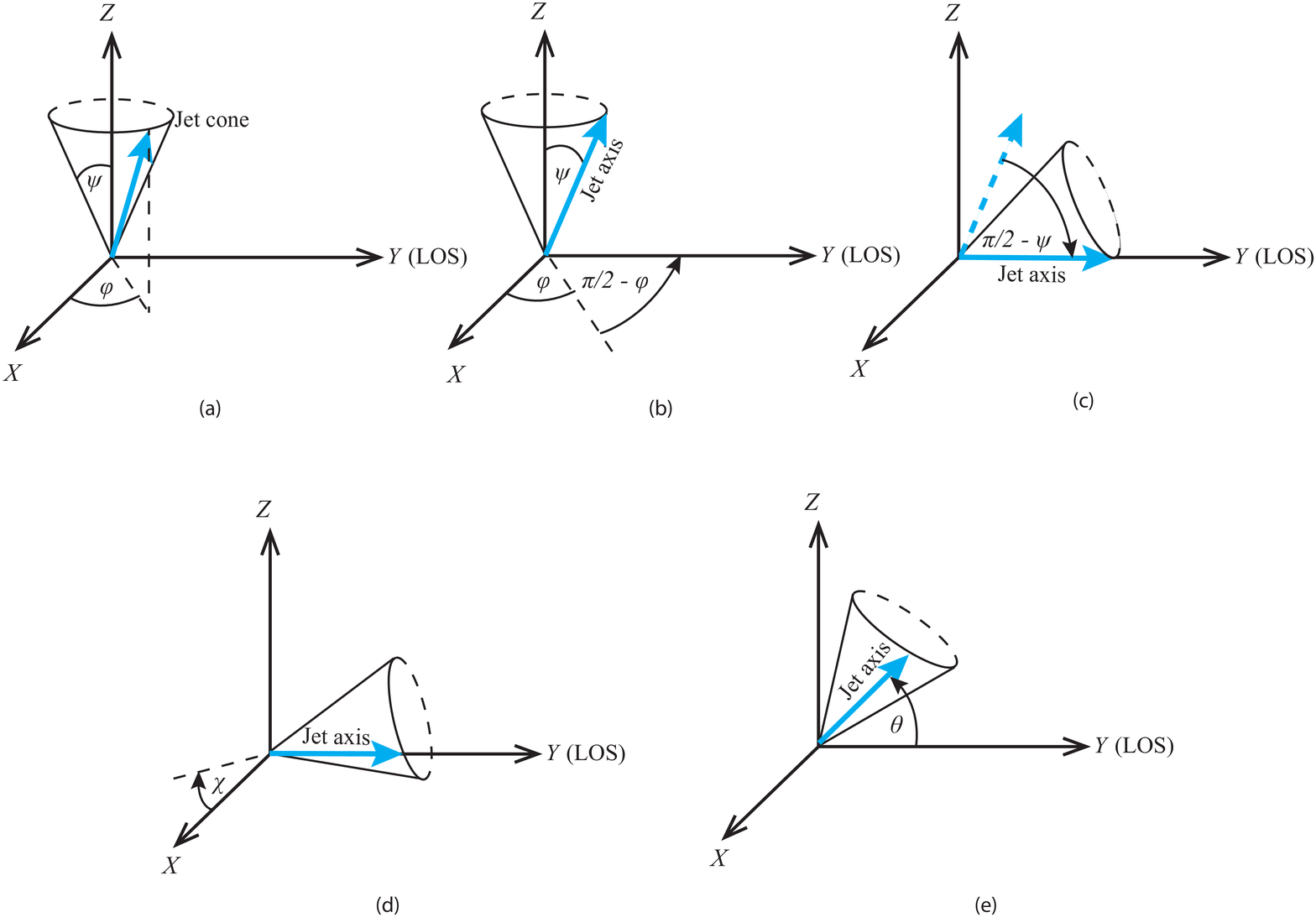}
\caption[Transformations associated with point rotations of the simulation data cube with respect to the synthetic image cube]{Transformations associated with point rotations of the simulation data cube with respect to the synthetic image cube Panel (a): Initially, the data cube and the image cube have the same orientation and are represented by the same coordinate system $XYZ$, which is that of the image cube. The jet axis and the jet cone are shown by a blue solid line and a black cone respectively. Far side of the cone top (ellipse) is shown in dashed line. Panel (b): A rotation about the $Z$-axis by the angle $\pi/2 - \phi$, the azimuthal angle of the precessing jet, brings the jet axis onto the $YZ$ plane. Panel (c): A rotation about the $X$-axis by the angle $\pi/2-\psi$, temporarily aligns the jet axis with the line of sight $Y$-axis. The direction of the jet before rotation is shown by the blue dashed line and after rotation by the solid blue line. Panel (d): A clockwise rotation about the LOS Y-axis by the angle $\chi$, orients the data cube to a specific azimuthal  direction about the jet axis. Panel (e): A rotation about the $X$-axis by the angle $\theta$, the angle between the jet and the line of sight, aligns the data cube to the desired line of sight. In each panel the precession cone is shown.}
\label{f:rot}
\end{figure*}

The visualisation software VISIT restricts the choice of the line of sight along any one axis of the image cube, which is used for ray-traced integrations of the surface brightness. Therefore, in order to prescribe a line of sight that is inclined at a specified angle 
$\theta$ to the jet direction and an azimuthal angle, $\chi$, about this direction, the following sequence of rotations of the data cube are required: 
Let $XYZ$ be the coordinates (shown in panel (a) of Fig.~\ref{f:rot}) associated with the synthetic image cube,  Let $Y$ be the direction of the line of sight. To begin with, the synthetic image cube and the simulation data cube have the same orientation. The jet direction is defined by the precession angle $\psi$ and the azimuthal angle $\phi$ defined with respect to the $X$-axis. The angle $\chi$ defines the azimuthal orientation of the data cube about the jet axis. As previously defined, $\theta$ is the angle between the jet axis and the line of sight $Y$-axis (see panel (e) of Fig.~\ref{f:rot}).
In Fig.~\ref{f:rot} angles are depicted by arcs and rotations are depicted by arcs with arrowheads. 
 
\begin{enumerate}
\item  We first rotate the simulation data cube anticlockwise about the $Z$-axis by an angle $\pi/2 - \phi$ (shown in panel (b)). 
This rotation brings the jet axis onto the $YZ$ plane. The rotation matrix for this rotation is given by 
\begin{eqnarray}
R^{(1)}_{Z,\phi}  &=& \begin{pmatrix}
 \cos(\pi/2 - \phi) & -\sin(\pi/2 - \phi) & 0 \\
\sin(\pi/2 - \phi) & \cos(\pi/2 - \phi) & 0 \\
0 & 0 & 1
\end{pmatrix} \nonumber \\
&=& \begin{pmatrix}
 \sin\phi & -\cos\phi & 0 \\
\cos\phi & \sin\phi & 0 \\
0 & 0 & 1
\end{pmatrix} 
\end{eqnarray}
\item We rotate the simulation data cube a second time clockwise about the $X$-axis by an angle 
$\pi/2 - \theta$ (shown in panel (c)). This rotation aligns the jet axis with the line of sight $Y$-axis. The matrix for this rotation is given by 
\begin{eqnarray}
R^{(2)}_{X, \psi} &=& \begin{pmatrix}
 1 & 0 & 0 \\
0 & \cos(\pi/2 - \psi) & \sin(\pi/2 - \psi) \\
0 & -\sin(\pi/2 - \psi) & \cos(\pi/2 - \psi)
\end{pmatrix} \nonumber \\
& = &\begin{pmatrix}
1 & 0 & 0 \\
0 & \sin\psi & \cos\psi \\
0 & -\cos\psi & \sin\psi
\end{pmatrix}  
\end{eqnarray}
\item In order to prescribe the azimuth of the viewing direction through the data cube, we rotate the data cube about the $Y$-axis by a 
clockwise angle $\chi$ (see panel (d)). With this rotation, the data cube rotates about the jet axis thereby changing the direction, about the jet axis, of the line of sight through the data cube. The matrix for this rotation is given by:
\begin{equation}
R^{(3)}_{Y,\chi} = \begin{pmatrix}
 \cos\chi & 0 & -\sin\chi \\
0 & 1  & 0 \\
\sin\chi & 0 & \cos\chi
\end{pmatrix}  
\end{equation}
\item Finally, we rotate the simulation data cube about the $X$-axis by the angle $\theta$ (shown in panel (e)). This rotation relocates the jet axis at the required angle $\theta$ with respect to the line of sight axis ($Y$). The rotation matrix associated with this rotation is given by
 \begin{equation}
  R^{(4)}_{X,\theta} = \begin{pmatrix}
 1 & 0 & 0 \\
0 & \cos\theta & -\sin\theta \\
0 & \sin\theta & \cos\theta 
\end{pmatrix}  
\end{equation}
\end{enumerate} 

The velocity of the fluid in the image cube $\textbf{v}'$ after the transformations described above is calculated from the velocity in the simulation data cube using the rotation matrix $R = R^{(4)}_{X,\theta} R^{(3)}_{Y,\chi} R^{(2)}_{X, \psi} R^{(1)}_{Z,\phi}$
\begin{equation}
\textbf{v}' = R \textbf{v}
\end{equation}
where $R$ is the combined transformation matrix. 

Let $s_1 = \sin \psi$, $s_2 = \sin \phi$, $s_3 = \sin \chi$, $s_4 = \sin \theta$, $c_1 = \cos \psi$,  $c_2 = \cos \phi$, $c_3 = \cos \chi$, and $c_4 = \cos\theta$. Then the transformation matrix $R$ is given by:
 \begin{eqnarray}
  R &=&  R^{(4)}_{X,\theta} R^{(3)}_{Y,\chi} R^{(2)}_{X, \psi} R^{(1)}_{Z, \phi} \nonumber \\
  &=&  \begin{pmatrix}
  c_3 s_1 + s_3 c_2 c_1  & c_3 c_1 + s_3 c_2 s_1 & -s_3 s_2 \\[6pt]
  c_4 s_2 c_1 - s_4 s_3 s_1  &  c_4 s_2 s_1 - s_4 s_3 c_1  & c_4 c_2 - s_4 s_2  \\
\phantom{{}+{}}+ s_4 c_3 c_1 c_2 & \phantom{{}+{}} + s_4 c_3 c_2 s_1  & & \\[6pt]
   s_4 s_2 c_1 + c_4 s_3 s_1 & s_4 s_2 s_1 + c_4 s_3 c_1  &  s_4 c_2 + c_4 s_2 \\
\phantom{{}+{}} - c_4 c_3 c_2c_1   &\phantom{{}+{}}-c_4 c_3 c_2 s_1  & & \\
\end{pmatrix} \nonumber 
\end{eqnarray}

\section{Projected angle between the jet axis and the precession axis on the sky}\label{A:proj}
\begin{figure*}
\centering
\includegraphics[width=\textwidth]{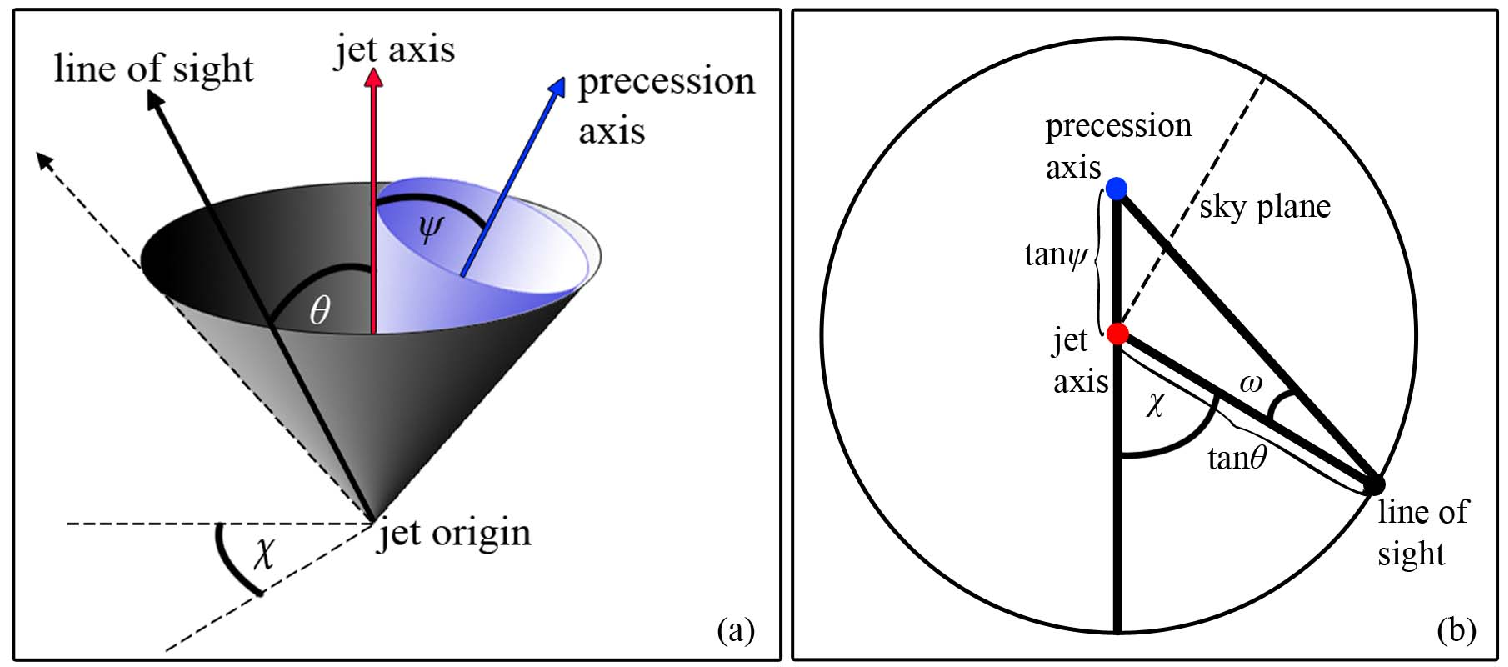}
\caption[Transformations associated with the rotations of points of the simulation data cube with respect to the synthetic image cube]{Panel (a) shows the line of sight cone (black cone) at an angle $\theta$ and the precession cone (blue cone) with precession angle $\psi$. The azimuth angle of line of sight $\chi$ is zero when the line of sight (black dashed arrow), jet axis (red arrow) and precession axis (blue arrow) all are on the same plane. At an arbitrary line of sight $\chi$ is shown with a black arrow. Panel (b) shows the perpendicular projection of the line of sight cone (circle) at a unit distance from the jet base. The line of sight, jet axis, and the precession axis are shown by black, red and blue points respectively. The projected angle on the sky plane (dashed line) between the jet axis and the precession axis is $\omega$. }
\label{f:pan}
\end{figure*}

Panel (a) of Fig.~\ref{f:pan} shows the line of sight cone (black cone) and the jet precession cone (blue cone) and the relative positions of the jet axis (red arrow), precession axis (blue arrow) and the line of sight (black arrow). The half cone angle of the line of sight cone is $\theta$. The azimuth angle of the line of sight, $\chi$, is measured from the direction perpendicular to the jet axis and coplanar with the jet axis and the precession axis. Panel (b) of Fig.~\ref{f:pan} shows the projection of the line of sight cone (circle) along the jet axis onto a plane at a unit distance from the jet origin. Therefore, the radius of the projected circle is $\tan\theta$ and the distance between the jet axis and precession axis is $\tan\psi$. Now, let the origin of the circle be (0,0). Then on the projected circle any arbitrary line of sight with $\chi$ is at ($\tan\theta \cos\chi, \tan\theta \sin\chi$) and the precession axis is at ($-\tan\psi, 0$). Using the \textit{cosine} rule, and solving for the projected angle $\omega$ between the jet axis and the precession axis on the sky plane (dashed line) we obtain 
\begin{equation}
\omega = \cos^{-1} \left (\frac{\tan\theta + \cos \chi \tan \psi}{\sqrt{\tan^2\theta + tan^2\psi + 2 \tan \theta \cos\chi \tan\psi}} \right ).
\end{equation}

\end{document}